\documentclass[conference]{IEEEtran}
\IEEEoverridecommandlockouts
\usepackage{amsmath,amssymb,amsfonts}
\usepackage{algorithmic}
\usepackage{graphicx}
\usepackage[hyphens]{url}
\usepackage{hyperref}
\usepackage[hyphenbreaks]{breakurl}
\usepackage{textcomp}
\usepackage{xcolor}
\def\BibTeX{{\rm B\kern-.05em{\sc i\kern-.025em b}\kern-.08em
    T\kern-.1667em\lower.7ex\hbox{E}\kern-.125emX}}
\usepackage[switch]{lineno}
\begin{document}
\pagenumbering{gobble}
\bstctlcite{IEEEexample:BSTcontrol} 

\title{Comparative Analysis of FPGA and GPU Performance for Machine Learning-Based \\ Track Reconstruction at LHCb}

\author{\IEEEauthorblockN{Fotis I. Giasemis}
\IEEEauthorblockA{\textit{LIP6, LPNHE} \\
\textit{Sorbonne Université}\\
\textit{CNRS, IN2P3}\\
Paris, France \\
Fotis.Giasemis@cern.ch}
\and
\IEEEauthorblockN{Vladimir Lončar}
\IEEEauthorblockA{\textit{Massachusetts Institute of Technology}\\
Cambridge, MA, USA \\
Vladimir.Loncar@cern.ch}
\and
\IEEEauthorblockN{Bertrand Granado}
\IEEEauthorblockA{\textit{LIP6} \\
\textit{Sorbonne Université}\\
\textit{CNRS}\\
Paris, France \\
Bertrand.Granado@lip6.fr}
\and
\IEEEauthorblockN{Vladimir Vava Gligorov}
\IEEEauthorblockA{\textit{LPNHE} \\
\textit{Sorbonne Université}\\
\textit{CNRS, IN2P3}\\
Paris, France \\
Vladimir.Gligorov@cern.ch}
}


\maketitle
\thispagestyle{plain} 
\pagestyle{plain} 

\begin{abstract}
In high-energy physics, the increasing luminosity and detector granularity at the Large Hadron Collider are driving the need for more efficient data processing solutions. Machine Learning has emerged as a promising tool for reconstructing charged particle tracks, due to its potentially linear computational scaling with detector hits. The recent implementation of a graph neural network-based track reconstruction pipeline in the first level trigger of the LHCb experiment on GPUs serves as a platform for comparative studies between computational architectures in the context of high-energy physics. This paper presents a novel comparison of the throughput of ML model inference between FPGAs and GPUs, focusing on the first step of the track reconstruction pipeline---an implementation of a multilayer perceptron. Using HLS4ML for FPGA deployment, we benchmark its performance against the GPU implementation and demonstrate the potential of FPGAs for high-throughput, low-latency inference without the need for an expertise in FPGA development and while consuming significantly less power.

\end{abstract}

\begin{IEEEkeywords}
FPGA, GPU, HLS4ML, machine learning, throughput, high energy physics
\end{IEEEkeywords}

\section{Introduction}

The LHCb Upgrade~I detector for Run~3~\cite{lhcb_collaboration_framework_2012} is a general purpose particle physics detector at the Large Hadron Collider (LHC). The Vertex Locator (VELO)~\cite{collaboration_lhcb_2013} is LHCb's innermost charged-particle tracking detector, surrounding the proton-proton interaction region and consisting of 52 planes of silicon pixels, each measuring $55\times55$ ${\mu m}^2$. 

The combined output from all the LHCb subdetectors can reach a data rate of up to 5~TB/s. To manage this volume before storage in Run~3, LHCb introduced Allen, a GPU-based online trigger system~\cite{lhcb_collaboration_lhcb_2020}, which reduces the data rate to 70-200~GB/s. Allen employs around 500 GPUs to partially reconstruct events, offering better selection efficiency than the FPGA-based hardware trigger used in LHCb during Run~1 and Run~2. Subsequently, the data is buffered for real-time alignment and calibration. A second, x86-based trigger then fully reconstructs the events and saves information containing physics signals of interest. The final processed data is stored at a rate of 10~GB/s.

In this environment, Allen and LHCb offer a platform for deploying neural network-based algorithms on GPUs, enabling high-throughput inference. However, GPUs are not the only suitable architecture for this application. FPGAs are also critical in high-energy physics, frequently utilized for tasks such as data acquisition, the processing involved in reading the data off the detectors, data compression, and high-speed data links. They offer the potential for better computational and energy efficiency, since they are configured and optimised for specifically one task. In addition, given the rising interest in machine learning within the high-energy physics community, it is essential to evaluate the suitability of FPGAs for offloading ML algorithms.

Indeed, ML inference on FPGAs has recently gained interest~\cite{nurvitadhi_can_2017,suda_throughput-optimized_2016}, and there have been various attempts in the context of high energy physics. Some ML techniques have been already deployed in the hardware part of the LHC trigger, including boosted decision trees for the inference of the momentum of muons in Level-1 of the CMS trigger~\cite{acosta_boosted_2017} and a convolutional neural network replacing a pattern-finding algorithm operating on hits in the detector in Level-0 of the ATLAS trigger~\cite{giagu_fast_2020} .

More specifically, in LHCb, FPGAs are used for the detector readout. With Upgrade~II of the LHCb detector planned in the 2030s~\cite{lhcb_collaboration_physics_2016}, it is therefore interesting to explore to what extent parts of the pattern recognition algorithms involved in the trigger, currently mostly classical but potentially incorporating more and more machine learning methods, can be moved ``closer'' to the detectors by performing them on the data acquisition FPGA boards, potentially improving the cost-effectiveness and energy efficiency of the experiment.

As the complexity of ML models used in high-level triggers for various experiments increases, it becomes crucial to study the fundamental components of these algorithms, such as the basic multilayer perceptron (MLP). Comparing the pros and cons of FPGAs with other processors in various contexts will provide valuable insights into their effectiveness and potential in high-energy physics applications. For this reason, in this work we focus on the first step of ETX4VELO, the Graph Neural Network (GNN)-based track reconstruction pipeline for the VELO, which involves an MLP. In addition, before the acceleration of the GNN is attempted, the workflow as well as the tools used for it have to be tried and optimized for a simpler model, and the MLP provides the ideal example.

For the FPGA deployment of the ML models, HLS4ML~\cite{duarte_fast_2018, fastml_team_hls4ml_2024} is used. The Python library is designed to be user-friendly, even for individuals without extensive experience in FPGA programming. This accessibility is particularly beneficial for the high energy physics community, where many researchers may not be experts in FPGA design. By lowering the barrier to entry, HSL4ML enables a broader range of scientists to leverage the benefits of FPGA acceleration in their work.

The framework consumes the model representation from various frameworks like Keras/TensorFlow~\cite{keras_developers_keras_nodate, abadi_tensorflow_2016} or PyTorch~\cite{paszke_pytorch_2019} and generates the code used by the High-Level Synthesis (HLS) tool in order to generate the Verilog/VHDL code, effectively hiding all the nuances of writing register-transfer level (RTL) code. It is designed for applications where low-latency and high-throughput implementations are critical. The board used for the implementation is a PYNQ-Z2 and the GPU is compared against the Alveo U50 and U250 data center accelerator cards.

In Section~\ref{pipeline}, we briefly present ETX4VELO, the GNN-based track-finding pipeline for the VELO, recently implemented in LHCb's first-level trigger, Allen. It is based on the Exa.TrkX collaboration's approach which interestingly exhibited a near-linear relationship between the throughput and the input event size~\cite{ju_performance_2021}, contrasting with the quadratic nature of conventional combinatorial algorithms. In the next section, Section~\ref{fpga-impl}, we present our workflow using HLS4ML for implementing and deploying the MLP model involved in the first step of the pipeline on an FPGA, its throughput, our estimate for the maximum throughput achievable on the Alveo FPGAs as well as its comparison with the GPU implementation.

\section{ETX4VELO Pipeline on GPU}
\label{pipeline}

The hits left by charged particles in the VELO detector can be represented as a graph, where hits from the same particle are connected to each other. The primary concept behind a GNN pipeline is to construct this graph of potential connections between hits in the detector, classify these connections as either connected or not, and then convert these connections into track objects that can be utilized by the rest of LHCb's real-time pipeline. Since a graph consisting of all possible hit connections would be prohibitively large, a significant challenge is to form the initial graph in a way that includes almost all genuine connections while excluding as many false connections as possible. In the ETX4VELO pipeline, this process involves an embedding MLP.

With the suitable parameters, the ETX4VELO pipeline has equivalent physics performance to that of LHCb's first-level trigger, Allen, while performing particularly well on reconstructing electrons, a challenging-to-reconstruct category of particles. For the full description of how the pipeline works and more details on each step, we refer the reader to~\cite{correia_graph_2024, correia_graph_2023}.

The ETX4VELO pipeline is implemented in C++/CUDA using the Allen framework to evaluate its computational performance. Computational throughput is measured using the Nvidia GeForce RTX 3090 card.

Here, we focus on the first step of the pipeline, which involves an embedding MLP. For the inference of the embedding network, the TensorRT (TRT)~\cite{tensorrt_developers_nvidia_2024} inference engine was utilized. The MLP model is quantized to INT8 precision using Nvidia's PyTorch-Quantization library~\cite{nvidia_developers_pytorch-quantization_nodate}, which is optimized for the TRT backend, and calibrated using 5000 events.

The current throughput of the pipeline on the GPU up to the embedding MLP, using the TensorRT implementation in INT8, is $820 \times 10^3$ events per second. The result quoted is from Allen's built-in throughput timer. It measures the number of events processed per second, for the series of algorithms specified. It excludes any initializations and/or optimizations of the algorithms before the main execution time. For more details on the GPU implementation see~\cite{correia_graph_2024}.

\section{FPGA Implementation}
\label{fpga-impl}

We explore the implementation of the ETX4VELO embedding MLP on FPGA hardware to compare its throughput against its GPU implementation. We utilize the HLS4ML library to convert the trained model to FPGA firmware and deploy it with AMD's open-source PYNQ project~\cite{amd_developers_pynq_nodate}. PYNQ offers a Jupyter-based framework with Python APIs for leveraging AMD adaptive computing platforms.

\subsection{Workflow}

Using HLS4ML, we convert the trained embedding MLP to HLS code, which is then synthesized to Verilog/VHDL, using Vivado HLS~\cite{amd_developers_vivado_nodate}. The workflow, for boards supported by the PYNQ project, involves the following steps. 

\subsubsection{Model Import} 
The trained PyTorch model is saved in PyTorch's native checkpoint format and is then imported. Since HLS4ML supports mainly Keras/TensorFlow, and has limited support for PyTorch, we had to use ReLU for the activations, and remove the LayerNorm layers, used for stabilizing and speeding up the training process, in order for the model to be processed by the HLS4ML library. The same modifications were performed on the GPU implementation.

\subsubsection{Model Configuration} 
The model parameters and FPGA target settings are configured in HLS4ML. For example, the optimization strategy is chosen by setting the \texttt{strategy} keyword to either ``Latency'' or ``Resource'' in order to guide the design in prioritizing latency or resource utilization, respectively. Here we choose the former. We define the precision of the inputs, outputs, weights and biases, and here we choose \texttt{ap\_fixed<16,6>}, where 6 is the number of bits representing the signed number above the binary point (i.e. the integer part), and 16 is the total number of bits.

\subsubsection{HLS Conversion} 
The model is converted to HLS code using HLS4ML. The model, along with the shape of the input data and the target FPGA is passed to the HLS4ML converter for PyTorch models which then parses and interprets the layers of the MLP, and then creates the HLS project. For the case under study, the MLP model is a fully-connected feed-forward neural network of 3-dimensional input, three hidden layers of 8 neurons each, ReLU activations, and a 3-dimensional output. The model is finally compiled with Vivado HLS. Version 2020.1 is used.

In HLS4ML, there are the concepts of the frontend and backend. The frontend is responsible for parsing the input neural network into an internal model graph, while the backend determines the type of output generated from this graph. Here we choose the VivadoAccelerator backend. The VivadoAccelerator backend of HLS4ML leverages the PYNQ software stack to deploy models on supported devices. For this backend, the I/O type, the hardware part that is being targeted, the clock period etc. has to be specified. The target FPGA boards for our implementations are the PYNQ-Z2 board, which contains a Xilinx Zynq-7020 FPGA, and the Alveo U50 and U250 featuring the UltraScale+ and XCU250 FPGAs, respectively. It is important to note that the PYNQ-Z2 board is designed for educational purposes, whereas the Alveo cards are significantly larger and intended for use in data centers. All three boards are supported by the PYNQ project.

\subsubsection{Synthesis and Implementation} 
The HLS code is synthesized to Verilog/VHDL. The model is finally ready to be synthesized with Vivado HLS. At this point, we can optionally perform the C simulation of the code, a process where the code is validated for errors and segmentation faults. The IP core is exported and the bitstream is saved.

\subsubsection{Deployment} 
The RTL implementation is deployed on the FPGA. We transfer the bitstream produced by Vivado, the hardware handoff file, used in building a platform for the target device, the driver and some data to the FPGA, and run the model using PYNQ Overlays. In the PYNQ project, programmable logic circuits are presented as hardware libraries called Overlays. The overlay can be accessed through a Python interface running in the Processing System (PS) of the FPGA, to allow the user to reconfigure the Programmable Logic (PL) of the FPGA. In HLS4ML, we create a custom neural network overlay, which sends and receives data via the AXI-Stream communication bus protocol. The target board is configured using the bitstream file generated by the VivadoAccelerator backend. Finally, using Python on the PS, we can create a NeuralNetworkOverlay object, which downloads the bitfile to the PL of the board. We also need to specify the shapes of our input and output data to allocate the necessary buffers for data transfer. The predict method sends the input data to the PL via the AXI-Stream and returns the output data.

\subsubsection{Computational Performance}
We benchmark the throughput of the model on the FPGA.  The inference can be timed using the built-in profiling tool of PYNQ Overlays. The current 16-bit implementation on the PYNQ-Z2 board achieves a throughput of approximately 1.2 million inferences per second. For an average LHCb event of 2200 VELO hits, using the same sample used in~\cite{correia_graph_2024}, the effective throughput comes out to 550 events per second.

\subsubsection{Model Precision}

In order to quantify the quality of the inference of the model on the FPGA, the various steps in the workflow have to be evaluated. The model in PyTorch has single precision (FP32) while in our chosen implementation 16 bits are used, and therefore there will be some loss of precision, unless techniques like QAT and/or profiling are used. For now these techniques are left for future work, and the evaluation is performed using the untuned quantization parameters. Also, since the model is not a classifier, the traditional evaluation methods, like the receiver operating characteristic (ROC) curve, are not suitable.

Using HLS4ML's method \texttt{predict} on the compiled HLS model of the MLP, we can get the predictions for the input array of approximately 200\,000 hits from the sample used in the GPU implementation. This is similar to doing the C simulation of the code, but the prediction results are more easily accessed.

For the \texttt{<16,6>} implementation 97\% of values are predicted within 10\% of correct values. Similarly to~\cite{correia_graph_2024}, when the precision of the embedding slightly decreases, the GNN is still able to maintain the physics performance of the pipeline almost unchanged.

Finally, the HLS4ML \texttt{predict} output is validated against the inference of the model on the hardware. The predictions on the PYNQ-Z2 card match perfectly the HLS4ML predictions on CPU.

\subsection{FPGA and GPU Comparison}

In order to compare with the INT8 implementation on the GPU, we use an Alveo U250 implementation of the network in 8-bit precision. From the Vivado estimates for the Alveo U250 card, the latency of the implementation is estimated to be at 85~ns. From this we can compute the effective theoretical throughput achievable to 11.8 million inferences per second. Given that the average number of hits is 2200, we get an effective throughput estimate of 5300 events per second. The utilization estimates are summarized in Table~\ref{tab:utilization}.

\begin{table}
    \centering
    \setlength{\tabcolsep}{2pt}
    \caption{Vivado synthesis report for resource utilization for the 8-bit implementation of the ETX4VELO embedding MLP on the Alveo U250 card.}
\begin{tabular}{c|c|c|c|c|c}
\hline \hline
Name & BRAM\_18K & DSP48E & FF & LUT & URAM \\ \hline
DSP & - & - & - & - & - \\
Expression & - & - & 40 & 1785 & - \\
FIFO & - & - & - & - & - \\
Instance & - & 0 & 410 & 6462 & - \\
Memory & - & - & - & - & - \\
Multiplexer & - & - & - & 149 & - \\
Register & - & - & 272 & - & - \\
Total & 0 & 0 & 722 & 8396 & 0 \\ \hline
Available & 5376 & 12\,288 & 3\,456\,000 & 1\,728\,000 & 1280 \\
Utilization (\%) & 0 & 0 & $\sim$0 & $\sim$0 & 0 \\ \hline \hline
\end{tabular}
    \label{tab:utilization}
\end{table}

We can extrapolate, based on the resource usage estimates from Vivado the number of available IPs we might be able to deploy on this high-end card. Based on this, we can quote the theoretically maximum throughput achievable.

Due to the implementation being on 8 bits, the DSPs were not allocated and thus the LUTs, being the most utilized resource, provide us with the maximum number of IPs we could launch on this platform. The Alveo U250 board has 1\,728\,000 LUTs, and the current 8-bit implementation of the ETX4VELO model on this card is using 8396 of them. Assuming, for simplicity, that the resource usage is not going to change dramatically as we pack more IPs on the same board, we should be able to launch at most $1\,728\,000 / 8396 \approx 205 $ IPs on this card. Hence, the maximum throughput achievable comes out to $205 \times 5300 = 1.1 \times 10^6$ events per second.

Finally, the same calculation is performed for the smaller Alveo U50 board. In this case, with a latency of 85~ns, and resources for approximately a maximum of 103 IPs, the throughput comes out to $550 \times 10^3$ events per second.


When the implementations with the specified flags are running, the GPU reaches its maximum power consumption of $350$~W. For the Alveo cards, we refer to the official specifications, which quote a maximum total power consumption of 75~W and 225~W for the U50 and U250, respectively, assuming the implementation will utilize the hardware close to its maximum capacity.

We can also compare the energy per event using the throughput and the power. Dividing the thermal design power of each device, in joules per second, by the throughput expressed in events per second, results in the energy cost of a single event. The GPU results in 430~$\mu$J per event, while the Alveo U50 and U250 result in 140 and 210~$\mu$J per event, respectively.

The prices of the accelerators are also considered. At the time of writing, the Alveo U50 is listed at 2965~USD on the official AMD website~\cite{amd_alveo_nodate-1}, while for the GPU, the launch price of 1499~USD~\cite{techpowerup_nvidia_2025} is used. The price of the Alveo U250 is not listed on the official website~\cite{amd_alveo_nodate}, so we estimate its current market price at approximately 10\,000~USD based on publicly available sources.

We now compare with the GPU throughput, which does not include memory transfers between the host and the device, since the data always reside on the device due to the Allen architecture. On one hand, the 8-bit implementation on the Alveo U250 is on par with the implementation on the GeForce RTX 3090, with the potential to slightly outperform it, while consuming just over 60\% of the power used by the GPU counterpart. However, it should be noted that the price of the U250 is roughly ten times the price of the GPU. On the other hand, the implementation on the Alveo U50 is slightly slower than the GPU counterpart, while the power usage is almost $5\times$ lower. Interestingly, the choice between FPGAs and GPUs involves a trade-off between their upfront cost and the long-term expense of power consumption over their operational lifetime. The results are summarized in Table~\ref{tab:fpga-gpu-comparison}.

\begin{table}
\centering
\setlength{\tabcolsep}{5pt}
\caption{Comparison of the embedding MLP throughput between the theoretical performance of the Alveo FPGA implementations and the GeForce RTX 3090 GPU implementation. For the Alveo implementations, \texttt{<a,b>} refers to \texttt{ap\_fixed<a,b>} precision. The power usage, the energy cost of the inference of a single event, and the price are also compared. The energy gain is given with respect to the GPU implementation. }
    \begin{tabular}{c|c|c|c}
        \hline \hline
        Accelerator & Alveo U50 & Alveo U250 & RTX 3090 \\
        Implementation & \texttt{<8,3>} & \texttt{<8,3>} & TRT INT8 \\
        \hline
        Throughput (Events/s $\times 10^{\rule{0pt}{1.5ex}6}$) & 0.55 & 1.10 & 0.82 \\
        Active Power Draw (W) & 75 & 230 & 350 \\
        Energy per Event ($\mu$J) & 140 & 210 & 430 \\
        Energy Gain & 3.1x & 2.0x & 1.0x \\ \hline
        Price (USD) & 3000 & $\sim$ 10\,000 & 1500 \\
        \hline \hline
    \end{tabular}
\label{tab:fpga-gpu-comparison}
\end{table}

We now proceed to compare the initial purchase cost of the accelerators with their operating costs, specifically focusing on electricity expenses. We will be using the cost of 125.2~EUR/MWh at the time of writing from the Swiss Federal Office of Energy website on electricity prices~\cite{swiss_federal_office_of_energy_sfoe_energy_nodate}. In order to simplify the calculation, we assume that the devices will be used at their maximum capacity throughout the entire year. The total hours will thus be 24~h $\times$ 365 = 8760~h. Therefore, running the GPU for one year at 350 W, would consume approximately 3.1 MWh, costing around 380 EUR. In contrast, the Alveo U50 would be about 3.1 times more cost-efficient, as shown in Table~\ref{tab:fpga-gpu-comparison}. Specifically, the Alveo U50 would cost 120 EUR to operate for a year, saving roughly 260 EUR annually compared to the GPU. Recovering, therefore, the 1500~USD price difference between the Alveo U50 and the GeForce RTX 3090---assuming, for the sake of simplicity, a USD/EUR exchange rate of 1:1---would take approximately 6 years.

Finally, for a more complete comparison, maintenance, upgrade, development, and optimization costs should also be considered; however, this is left for future work.

\section{Conclusions}
\label{conclusions}

This work presents a detailed comparison of machine learning inference on FPGAs and GPUs in the context of future high-energy physics experiments, where scalability, power efficiency, and computational performance will be increasingly critical. These findings underscore the potential of FPGAs as viable alternatives for high-throughput applications in particle physics, especially when energy is an important consideration. The combination of HLS4ML's ease of use and the inherent advantages of FPGAs makes this approach a compelling choice for researchers aiming to implement ML models in hardware without deep expertise in FPGA development.

In addition to underlining the FPGA's strengths in terms of energy efficiency and throughput, this work also highlights areas for future optimization, such as quantization aware training in order to retain the required physics performance, or computational performance optimization using HLS directives. These optimizations could further unlock the full potential of FPGAs, making them even more suitable for the high-performance demands of real-time data processing at the LHC.

\section*{Acknowledgments}
This work is part of the SMARTHEP network and it is funded by the European Union’s Horizon 2020 research and innovation programme, call H2020-MSCA-ITN-2020, under Grant Agreement n. 956086. It is also supported by the ANR-BMBF project ANN4EUROPE ANR-21-FAI1-0011, in collaboration with the Frankfurt Institute for Advanced Studies (FIAS).

Our thanks extend to Anthony Correia, Nabil Garroum, and the ANN4Europe group at FIAS, led by Ivan Kisel, for their invaluable discussions. We are also grateful to Maarten van Veghel and the LHCb Real-Time Analysis project for their support, insightful discussions, and for reviewing an early draft of this manuscript. Special thanks go to Roel Aaij and the LHCb engineering teams for assisting with the setup of environment dependencies for model inference in Allen, and to the LHCb computing and simulation teams for generating the simulated samples used to benchmark the algorithm's performance.

\bibliographystyle{IEEEtran}
\bibliography{IEEEabrv,references}

\end{document}